\documentclass[
 reprint,
 amsmath,amssymb,
 aps,showkeys,superscriptaddress,pra,floatfix
]{revtex4-2}

\usepackage{graphicx}
\usepackage{dcolumn}
\usepackage{bm}

\begin{document}


\title{Investigating forward–backward asymmetry in D-meson production and anisotropic flow in p–Pb collisions at the LHC} 



\author{Siyu Tang}
    \email[Correspondence email address: ]{tsy@wtu.edu.cn}
    \affiliation{School of Electronic and Electrical Engineering, Wuhan Textile University, Wuhan 430200, China}
	\affiliation{Shanghai Research Center for Theoretical Nuclear Physics, NSFC and Fudan University, Shanghai 200438, China}
\author{Chao Zhang}
    \affiliation{School of Physics and Mechanics, Wuhan University of Technology, Wuhan, 430070, China}
	\affiliation{Shanghai Research Center for Theoretical Nuclear Physics, NSFC and Fudan University, Shanghai 200438, China}
\author{Liang Zheng}
    \affiliation{School of Mathematics and Physics, China University of Geosciences (Wuhan), Wuhan 430074, China}
	\affiliation{Shanghai Research Center for Theoretical Nuclear Physics, NSFC and Fudan University, Shanghai 200438, China}
\author{Renzhuo Wan}
    \email[Correspondence email address: ]{wanrz@wtu.edu.cn}
    \affiliation{School of Electronic and Electrical Engineering, Wuhan Textile University, Wuhan 430200, China}
\author{Zi-Wei Lin}
    \email[Correspondence email address: ]{linz@ecu.edu} 
    \affiliation{Department of Physics, East Carolina University, Greenville, NC 27858, USA}
\author{Guo-Liang Ma}
    \email[Correspondence email address: ]{glma@fudan.edu.cn}
	\affiliation{Key Laboratory of Nuclear Physics and Ion-beam Application (MOE), Institute of Modern Physics, Fudan University, Shanghai 200433, China}
	\affiliation{Shanghai Research Center for Theoretical Nuclear Physics, NSFC and Fudan University, Shanghai 200438, China}
\date{\today} 


\date{\today}

\begin{abstract}
We investigate the forward--backward asymmetry in the production and elliptic flow of prompt $D^0$ mesons in proton--lead (p--Pb) collisions at $\sqrt{s_{\mathrm{NN}}}=8.16$~TeV using the heavy-flavor improved string-melting version of the AMPT model. The model calculations provide a simultaneous description of nuclear modification factor $R_{\mathrm{pPb}}$ and $v_2$ in forward and backward rapidities. We find that the observed asymmetry arises from the interplay of initial-state cold nuclear matter effects and final-state partonic interactions, with the competition between coalescence and fragmentation playing a critical role in shaping the transverse momentum and rapidity dependence of both observables. This work suggests that a partonic medium is formed in high-multiplicity p-–Pb collisions at LHC energies.
\end{abstract}

\keywords{small collision systems, heavy-flavor hadrons, forward--backward asymmetry, parton transport, LHC}
\maketitle

\section{Introduction}      
\label{sec:introduction}

High-energy heavy-ion collisions at the Relativistic Heavy Ion Collider (RHIC) and the Large Hadron Collider (LHC) provide a unique platform for investigating nuclear matter under extreme conditions of temperature and density~\cite{Gyulassy:2004zy,STAR:2005gfr,PHENIX:2004vcz,Muller:2012zq,ALICE:2022wpn,CMS:2024krd,PHOBOS:2004zne,BRAHMS:2004adc}. In these collisions, a deconfined state of matter known as the quark-gluon plasma (QGP) is created~\cite{Shuryak:1978ij,Shuryak:1980tp}, where quarks and gluons are no longer confined within individual hadrons but instead form a hot and dense medium. The QGP behaves as a near-perfect fluid with an extremely low shear viscosity to entropy density ratio~\cite{Schaefer:2009}, as inferred from the large azimuthal anisotropy (collective flow) of produced particles~\cite{Heinz:2013th}. In addition, a strong suppression of high-transverse-momentum hadrons (jet quenching) is observed, which arises from parton energy loss in the dense medium~\cite{Qin:2015}. Heavy quarks (charm and bottom) are particularly powerful probes of this medium. Due to their large masses, they are produced in hard scatterings at the very early stage of the collision, before the QGP forms, and then traverse the entire evolution of the expanding system~\cite{Apolinario:2022}. Various measurements of heavy-flavor hadron production and collective flow thus shed light on the transport properties of the QGP and the mechanisms of heavy-quark hadronization in a hot and dense environment~\cite{ALICE:2022D0,ALICE:2022Ds,ALICE:2023LcD0,ALICE:2022BeautyD0,CMS:2021D0D0bar,ALICE:2023nonpromptD0,ALICE:2021DmesonFlow}.

In contrast to nucleus--nucleus collisions, proton--nucleus (p--Pb) collisions have long been regarded as an essential baseline for heavy-ion collisions at the LHC. Studies of heavy-flavor hadron production in p--Pb collisions are expected to isolate cold nuclear matter (CNM) effects~\cite{Salgado:2011wc}, such as nuclear shadowing and initial-state transverse momentum broadening, and thereby provide a crucial reference for disentangling initial-state modifications from hot QGP effects in larger systems. The ALICE collaboration has measured open charm production at midrapidity at $\sqrt{s_{\mathrm{NN}}}=5.02$~TeV~\cite{ALICE:2019fhe}, finding that the nuclear modification factor $R_{\mathrm{pPb}}$ of prompt $D$ mesons is consistent with unity within uncertainties, with a modest enhancement at low $p_{\mathrm{T}}$. The LHCb collaboration has measured prompt $D^0$ meson production at both forward and backward rapidities in p--Pb collisions at $\sqrt{s_{\mathrm{NN}}}=5.02$ and $8.16$~TeV~\cite{LHCb:2017yua,LHCb:2022dmh}, revealing a significant forward--backward asymmetry in $R_{\mathrm{pPb}}$. In the forward region, which probes small Bjorken-$x$ partons in the lead nucleus, a clear suppression of $R_{\mathrm{pPb}}$ is observed. In the backward region, corresponding to moderate $x$, $R_{\mathrm{pPb}}$ is close to unity or slightly enhanced~\cite{LHCb:2017yua}. Whether this asymmetry can be fully accounted for by initial-state cold nuclear matter effects alone, or whether final-state partonic interactions and the details of heavy-quark hadronization also contribute, remains an open question that calls for further theoretical investigation.

In recent years, collective flow phenomena for heavy flavors have also been observed in small collision systems~\cite{CMS:2012qk,ALICE:2012eyl,ATLAS:2012cix,ALICE:2023gyf,ALICE:2024vzv,ALICE:2025bwp,ATLAS:2023rbh,CMS:2025kzg,CMS:2023iam}. The CMS collaboration reported the first measurement of elliptic flow for prompt $D^{0}$ mesons in p--Pb collisions, finding a sizable $v_{2}$ that is nevertheless smaller than that of light-flavor hadrons~\cite{CMS:2018loe}. The ALICE collaboration has further measured significant $v_{2}$ signals for electrons and muons from heavy-flavor decays at mid, forward, and backward rapidities~\cite{ALICE:2018gyx,ALICE:2022ruh}. These observations indicate that heavy quarks participate in the collective expansion of the small collision system, yet the underlying mechanism remains under debate, particularly given the forward--backward asymmetry observed in the measured $v_{2}$. In hydrodynamics-based models, a sizable $v_{2}$ is naturally generated from strong interactions between charm quarks and the expanding medium. However, such interactions would also cause significant suppression of high-$p_{\mathrm{T}}$ $D$ meson yields, which contradicts the observed unity $R_{\mathrm{pPb}}$~\cite{Du:2018wsj,Cao:2020wlm,Xu:2015iha,Zhao:2020wcd}. Calculations based on initial-state effects, such as those within the Color Glass Condensate framework~\cite{Zhang:2019dth,Zhang:2020ayy}, can reproduce the heavy-flavor $v_{2}$ at midrapidity but overestimate the data at forward rapidities~\cite{ALICE:2022ruh}. In addition, the A Multi-Phase Transport (AMPT) model, which incorporates both initial-state effects and final-state interactions, is also widely used to investigate collective flow and particle production in small collision systems~\cite{Bzdak:2014dia,Bozek:2015swa,Tang:2023wcd,Tang:2024kot}. The studies with AMPT model have demonstrated that the partonic escape mechanism plays an essential role in generating azimuthal anisotropy in small collision systems~\cite{He:2015hfa}. The heavy-flavor improved string-melting version of AMPT model has been shown to provide a reasonable description of $R_{\mathrm{pPb}}$ and $v_2$ of $D^0$ meson in p--Pb collisions at mid-rapidity~\cite{Zhang:2022fum}, highlighting the importance of both parton interactions and the Cronin effect. A recent AMPT study has systematically investigated the forward and backward rapidity dependence of $D^0$ meson $R_{\mathrm{pPb}}$ in p--Pb collisions at $\sqrt{s_{\mathrm{NN}}}=5.02$~TeV, demonstrating that a consistent description of the rapidity-dependent nuclear modification requires a careful calibration of the Cronin broadening strength~\cite{Zhang:2024zga}. However, a unified description of $R_{\mathrm{pPb}}$ and $v_2$ of heavy flavour in the forward and backward rapidities is still lacking at the higher energy of 8.16~TeV, and the physical mechanisms driving their forward--backward asymmetry remain to be clarified. Understanding these mechanisms would be particularly valuable for the small-$x$ physics in the forward region. Therefore, a comprehensive study of $D^0$ meson production and $v_2$ in both forward and backward rapidities is required.

In this work, we employ the heavy-flavor improved string-melting version of the AMPT model to perform a systematic study of prompt $D^0$ meson production and elliptic flow $v_2$ in p--Pb collisions at $\sqrt{s_{\mathrm{NN}}}=8.16$~TeV. By comparing theoretical calculations with experimental data from the LHCb and ALICE collaborations, we investigate the forward--backward asymmetries observed in both the nuclear modification factor $R_{\mathrm{pPb}}$ and the azimuthal anisotropy $v_2$. The rest of this paper is organized as follows. In Sec.~\ref{sec:model} we briefly describe the heavy-flavor improved AMPT model and the two-particle correlation method used for flow extraction. Section~\ref{sec:results} presents the calculated $p_{\mathrm{T}}$ spectra, $R_{\mathrm{pPb}}$, and $v_2$ of prompt $D^0$ mesons in forward and backward rapidities, examining their sensitivity to various physical mechanisms. A summary is given in Sec.~\ref{sec:summary}.

\section{The AMPT model and Methodology}
\label{sec:model}
\subsection{A multiphase transport mode}
The AMPT model is a hybrid approach that simulates the full evolution of relativistic heavy-ion collisions in four stages: initial parton production, parton cascade, hadronization, and hadronic rescattering~\cite{Lin:2004en,Lin:2021mdn}. In this work, we employ the string-melting version of the AMPT model with improved treatments for heavy flavor~\cite{Zheng:2019alz}, as detailed in the following stages. 

In the string-melting version, the initial parton distribution is provided by the HIJING event generator~\cite{Gyulassy:1994ew}. For charm quarks, we extract them directly from the HIJING initial conditions, which include leading-order pair production $g+g\rightarrow c+\bar{c}$, $q+\bar{q}\rightarrow c+\bar{c}$, and gluon splitting $g\rightarrow c+\bar{c}$ via a parton shower. Nuclear shadowing effects are implemented by multiplying the free-nucleon parton distribution functions with a spatially dependent shadowing factor for the lead nucleus~\cite{Zhang:2021vvp}. To account for the Cronin effect, the initial transverse momentum of charm quarks is broadened by a Gaussian smearing with a width $\delta$. Given the small difference in beam rapidity between the two energies, we adopt the $\delta(y)$ parameterization from the 5.02 TeV analysis~\cite{Zhang:2024zga}, where it was determined via a fit to the $R_{\mathrm{pPb}}$ data. For reference, the typical $\delta$ values in the forward (2.5 $< y <$ 4.0) and backward (-4.0 $< y <$ -2.5) rapidity ranges are approximately 4.2 and 1.8, respectively. These charm quarks are then transported directly to the parton cascade stage, without undergoing the string melting process applied to the initial light flavor hadrons.

The parton cascade stage is implemented using Zhang's Parton Cascade (ZPC) model, where partons undergo two-body elastic scatterings with a cross section $\sigma = 9\pi\alpha_s^2/(2\mu^2)$, with $\alpha_s = 0.33$~\cite{Zhang:1997ej}. A key refinement for heavy flavor is the use of separate scattering cross sections, achieved by adjusting the screening mass $\mu$ in the ZPC framework. Light quarks are assigned $\sigma_{\mathrm{LQ}} = 0.5$~mb, while processes involving heavy quarks are assigned a larger cross section $\sigma_{\mathrm{HQ}} = 1.5$~mb. This distinction accounts for the difference in interaction strength between charm quarks and light partons, and has been shown to be essential for describing the $D^0$ meson $v_2$ in high-multiplicity p--Pb collisions~\cite{Zhang:2022fum,Zhang:2024zga}.

After the parton cascade, hadronization is performed via a two-component mechanism that accounts for the different phase-space regions of charm quarks. Charm quarks coalesce with nearby light partons more frequently at lower $p_{\mathrm{T}}$, while high-$p_{\mathrm{T}}$ charm quarks tend to fragment independently. To implement this competition, we employ a set of criteria based on the relative distance and invariant mass of the the coalescing 2 or 3 partons in their rest frame. Specifically, a charm quark $c$ and a light antiquark $\bar{q}$ (or a diquark) are considered suitable for coalescence if they satisfy
\begin{equation}
d < p_r , \qquad m_{\mathrm{inv}} < \sum m_Q + p_m \left(m_H - \sum m_Q\right),
\end{equation}
where $d$ and $m_{\mathrm{inv}}$ denote the relative distance and invariant mass, $m_Q$ and $m_H$ are the masses of the charm quark and the resulting $D$ meson, respectively. The parameters $p_r = 0.9$~fm and $p_m = 0.5$ were determined by fitting the $D^0$ meson $p_{\mathrm{T}}$ spectrum in $pp$ collisions~\cite{Zhang:2024zga}. Charm quarks that do not satisfy these conditions undergo independent fragmentation using the Peterson fragmentation function~\cite{Peterson:1982ak}
\begin{equation}
f(z) \propto \frac{1}{z\left(1 - \frac{1}{z} - \frac{\epsilon_Q}{1-z}\right)^2},
\end{equation}
where $z$ is the fraction of the charm quark's energy carried by the $D$ meson, and $\epsilon_c = 0.05$ is used for charm quarks. The fragmentation is implemented via PYTHIA~\cite{Sjostrand:1994kzr}, which also handles the formation of the accompanying light hadrons. In both channels, the hadronization produces not only the ground-state $D$ mesons but also their excited partners, such as the vector $D^*$ mesons and the charmed baryons ($\Lambda_c$, $\Sigma_c$, $\Xi_c$). The $D^*$ mesons subsequently decay into $D^0$ during the hadronic cascade stage. This two-component hadronization scheme has been demonstrated to be essential for describing the $D^0$ meson production and $v_2$ in p--Pb collisions at 5.02~TeV~\cite{Zhang:2022fum,Zhang:2024zga}, and we adopt the same framework to study the rapidity dependence of $D^0$ observables at 8.16~TeV.

Finally, the produced hadrons are propagated through the hadronic cascade stage using the ART model~\cite{Li:1995pra}, which accounts for rescattering among hadrons. In this work, we focus on prompt $D^0$ mesons, thus the contribution from bottom-quark decays is not included. The model calculations are performed for minimum-bias and high-multiplicity p--Pb collisions at $\sqrt{s_{\mathrm{NN}}}=8.16$~TeV, with the same parameter set as used in our previous work~\cite{Zhang:2024zga}.

\subsection{Two-particle correlation and nonflow subtraction}
To investigate the anisotropic flow of prompt $D^0$ mesons, we employ the two-particle correlation method, which has been widely used in small collision systems. In this approach, the correlation between trigger particles (prompt $D^0$ mesons) and associated particles (charged particles at midrapidity) is constructed as a function of the azimuthal angle difference $\Delta\phi$ and pseudorapidity difference $\Delta\eta$. The associated yield per trigger particle is defined as:
\begin{equation}
Y(\Delta\eta,\Delta\phi) = \frac{1}{N_{\text{trig}}} \frac{\mathrm{d}^2 N_{\text{assoc}}}{\mathrm{d}\Delta\eta \mathrm{d}\Delta\phi} = \frac{S(\Delta\eta,\Delta\phi)}{B(\Delta\eta,\Delta\phi)},
\end{equation}
where $N_{\text{trig}}$ is the number of trigger particles in a given multiplicity and $p_{\mathrm{T}}$ interval. The signal distribution $S(\Delta\eta,\Delta\phi)$ is obtained by correlating trigger particles with associated particles from the same event, while the background distribution $B(\Delta\eta,\Delta\phi)$ is constructed by mixing trigger particles from one event with associated particles from other events.

In high-multiplicity events, the correlation function distribution contains contributions from both collective flow and nonflow effects, such as jet correlations and resonance decays. To suppress these nonflow contributions, a two-step subtraction procedure is widely adopted in experimental analyses. The first step is to apply a pseudorapidity gap of $|\Delta\eta| > 1.0$ to suppress short-range nonflow correlations. The second step subtracts long-range jet-like correlations using low-multiplicity events as a reference, where the per-trigger yield in high-multiplicity events is subtracted by the scaled per-trigger yield obtained in low-multiplicity events. The scale factor is defined as the ratio of the corresponding yields in the away side in high-multiplicity collisions to those in low-multiplicity collisions, assuming that the shape of dijets are identical for both collision types~\cite{ALICE:2015lpx,ALICE:2017smo}. After subtraction, the residual correlation distribution is projected onto $\Delta\phi$ and fitted with a Fourier series:
\begin{equation}
\begin{aligned}
C^{\mathrm{HM}}(\Delta\varphi) - F\cdot C^{\mathrm{LM}}(\Delta\varphi) = a_{0} + 2 \sum_{n=1}^{\infty}a_{n}\mathrm{cos}[n(\Delta\varphi)],
\end{aligned}
\label{eq: peripheral subtraction}
\end{equation}
where $C(\Delta\varphi)=\frac{1}{N_{\text{trig}}} \frac{\mathrm{d} N_{\text{assoc}}}{\mathrm{d}\Delta\phi}$ is the projected correlation functions for high-multiplicity and low-multiplicity events, and $F$ is the scale factor. The second-order Fourier coefficient $a_2$ is used to extract the single-particle elliptic flow of $D^0$ mesons, $v_2^{D^0}$, by factoring out the reference flow of associated particles via:
\begin{equation}
v_2^{D^0} = \frac{a_2}{\sqrt{a_2^{\text{ref}}}},
\end{equation}
where $a_2^{\text{ref}}$ is the second-order Fourier coefficient from correlations among associated particles (e.g., charged particles at midrapidity). This two-particle correlation method, combined with the low-multiplicity subtraction, has been successfully applied to study the elliptic flow in p--Pb collisions~\cite{ALICE:2022ruh}. Adopting the same framework in this work allows for a direct comparison between our theoretical calculations and the experimental measurements.

\section{Results and Discussion}
\label{sec:results}
\subsection{Prompt $D^0$ meson production and $R_{\mathrm{pPb}}$}
\begin{figure*}[!hbt]
\begin{center}
\includegraphics[width=2.\columnwidth]{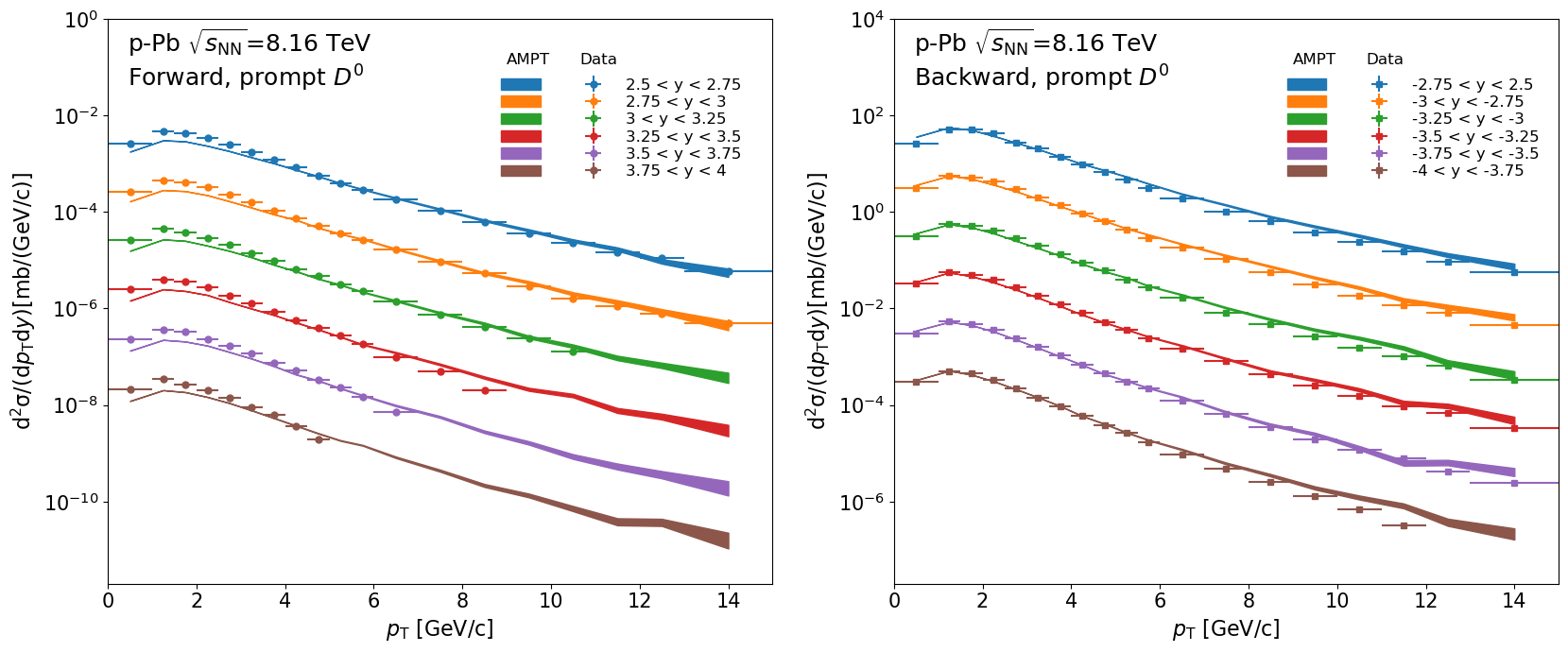}
\caption{(Color online) Double-differential cross-sections of prompt $D^0$ mesons in p--Pb collisions in the forward (left) and backward (right) rapidity regions. The data points are from the LHCb Collaboration~\cite{LHCb:2022dmh}. The lines are the AMPT model results.}
\label{Fig: D0 pt Spectrum}
\end{center}
\end{figure*}
We first study the $p_{\mathrm{T}}$ spectrum of prompt $D^0$ mesons with the improved AMPT model. Figure~\ref{Fig: D0 pt Spectrum} presents the double-differential production cross sections of prompt $D^0$ mesons as a function of $p_{\mathrm{T}}$ in p--Pb collisions at $\sqrt{s_{\mathrm{NN}}} = 8.16$ TeV for forward ($2.5 < y < 4.0$, proton-going side) and backward ($-4.0 < y < -2.5$, lead-going side) rapidity intervals, compared with the LHCb experimental data~\cite{LHCb:2022dmh}. A clear asymmetry between forward and backward rapidities is observed, with the yield in the forward region systematically lower than that in the backward region at the same $p_{\mathrm{T}}$. The AMPT model generally reproduces the measured $p_{\mathrm{T}}$ spectra, while it slightly underestimates the experimental data in the forward rapidity region for $p_{\mathrm{T}} < 2$ GeV/$c$, particularly in the more forward bins. This discrepancy aligns with earlier observations at 5.02 TeV~\cite{Zhang:2024zga}, suggesting a possible underestimation of the production mechanism in the very forward region within the current model setup, which may be attributed to uncertainties in the gluon nuclear shadowing parametrization~\cite{Helenius:2012wd,Eskola:2021nhw,Duwentaster:2022kpv} and the treatment of initial-state parton dynamics in asymmetric collision systems.

\begin{figure}[!hbt]
\begin{center}
\includegraphics[width=1.\columnwidth]{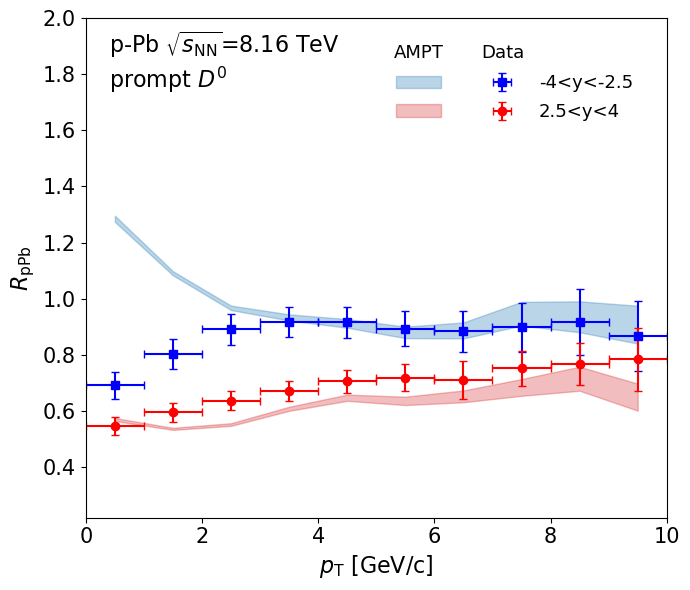}
\caption{(Color online) Nuclear modification factor as a function of $p_T$ for prompt $D^0$ mesons in the forward and backward regions. The data points are from the LHCb Collaboration~\cite{LHCb:2022dmh}. The lines are the AMPT model results.}
\label{Fig2_RpPb}
\end{center}
\end{figure}

We then investigate the nuclear modification factor $R_{\mathrm{pPb}}$ of prompt $D^0$ mesons. This observable is defined as the ratio of the double-differential production cross section of $D^0$ mesons in p--Pb collisions to that in $pp$ collisions at the same rapidity and collision energy, scaled by the atomic mass number of the lead nucleus:
\begin{equation}
R_{\mathrm{pPb}} = \frac{1}{A} \frac{d^2\sigma_{\mathrm{pPb}}/dp_{\mathrm{T}}dy}{d^2\sigma_{pp}/dp_{\mathrm{T}}dy},
\end{equation}
where $A = 208$ is the mass number of the lead nucleus. The $R_{\mathrm{pPb}}$ of $D^0$ mesons is sensitive to both cold nuclear matter effects, including nuclear shadowing and the Cronin effect, and the energy loss arising from parton scatterings during the evolution. Figure~\ref{Fig2_RpPb} presents the $R_{\mathrm{pPb}}$ of prompt $D^0$ mesons as a function of $p_{\mathrm{T}}$ in p--Pb collisions at $\sqrt{s_{\mathrm{NN}}} = 8.16$ TeV, for forward and backward rapidity intervals, compared with LHCb data~\cite{LHCb:2022dmh}. The AMPT model, using the $\delta$ value obtained from the fit to the $R_{\mathrm{pPb}}$ data across rapidities, generally reproduces the experimental measurements in both regions. In the backward rapidity and for $p_{\mathrm{T}} < 2$ GeV/$c$, the model calculation overestimates the data, a tendency also reported in earlier studies at 5.02 TeV~\cite{Zhang:2024zga}. A clear rapidity asymmetry is observed, with $R_{\mathrm{pPb}}$ exhibiting a suppression trend in the forward region and remaining close to unity in the backward region. This pattern is consistent with the expected influence of nuclear shadowing, which suppresses parton distributions at small Bjorken-$x$ (forward) and exhibits an anti-shadowing behavior at moderate $x$ (backward)~\cite{LHCb:2017yua}. 

\begin{figure*}[!hbt]
\begin{center}
\includegraphics[width=2.\columnwidth]{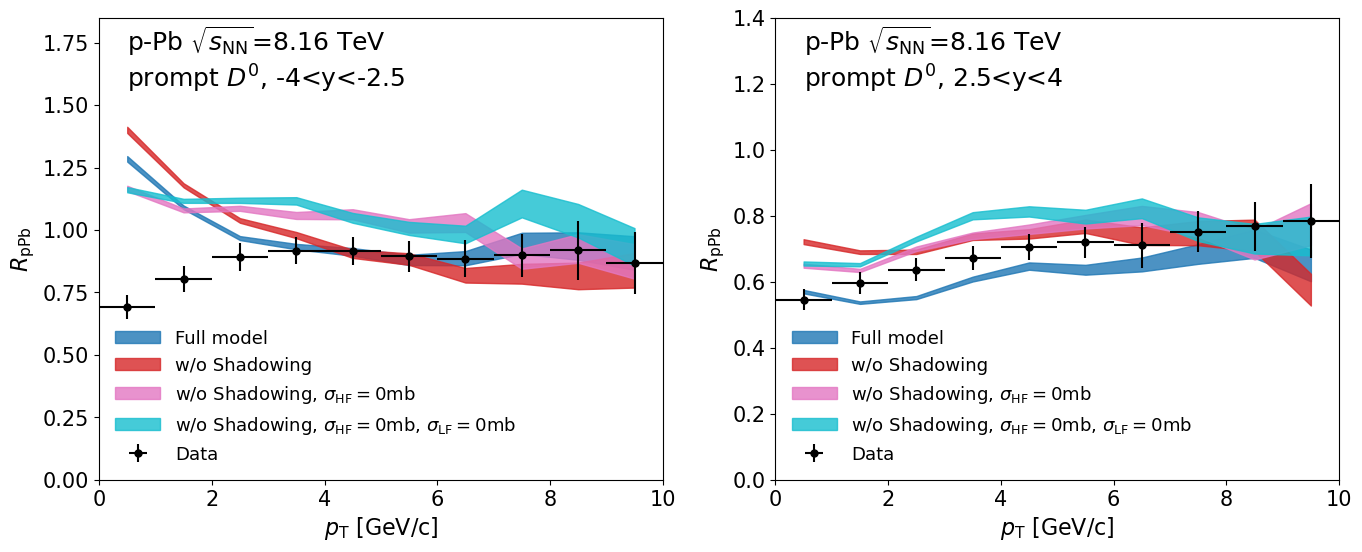}
\caption{(Color online) Nuclear modification factor as a function of $p_T$ for prompt $D^0$ mesons in the backward (left) and forward (right) regions. The lines are the AMPT model results with different model configurations, the data points are from the LHCb Collaboration~\cite{LHCb:2022dmh}.}
\label{Fig4_D0RpPb}
\end{center}
\end{figure*}

To further disentangle the contributions of different cold and hot nuclear matter effects to the observed nuclear modification, the $R_{\mathrm{pPb}}$ of $D^0$ mesons under various configurations of the AMPT model are calculated in both backward and forward rapidity intervals, as shown in Fig.~\ref{Fig4_D0RpPb}. The nuclear shadowing effect is observed to produce distinct modifications in the two rapidity regions. In the forward region, where small-$x$ partons are probed, the removal of shadowing enhances $R_{\mathrm{pPb}}$ across the entire $p_{\mathrm{T}}$ range, reflecting the release of previously suppressed parton densities. In the backward region, corresponding to moderate $x$, the effect exhibits a clear $p_{\mathrm{T}}$ dependence, where $R_{\mathrm{pPb}}$ increases at low $p_{\mathrm{T}}$ ($<4$ GeV/$c$) but slightly decreases at higher $p_{\mathrm{T}}$. This $p_{\mathrm{T}}$-dependent behavior can be attributed to the interplay between anti-shadowing at moderate $x$ and the Cronin effect, which redistributes partons from low to intermediate $p_{\mathrm{T}}$ via multiple scattering. These observations are consistent with the expected $x$-dependence of nuclear parton distribution modifications and provide a more differential understanding of the shadowing effect beyond the rapidity-integrated picture presented in earlier studies~\cite{Zhang:2024zga}. Furthermore, setting the heavy quark scattering cross section to zero ($\sigma_{\mathrm{HF}} = 0$) leads to a significant suppression of $R_{\mathrm{pPb}}$ at low $p_{\mathrm{T}}$ ($< 2$ GeV/$c$) and an enhancement at higher $p_{\mathrm{T}}$ in both rapidity regions. This behavior reflects the competition between collisional energy loss and transverse momentum redistribution, and the stronger enhancement of $R_{\mathrm{pPb}}$ in the backward region is consistent with the higher parton density on the lead-going side. Subsequently disabling light quark scatterings ($\sigma_{\mathrm{LF}} = 0$) results in a modest overall increase in $R_{\mathrm{pPb}}$ with similar magnitude in both rapidity regions, indicating that interactions with light partons also contribute to the modification but exhibit weaker rapidity dependence. 

\begin{figure}[!hbt]
\begin{center}
\includegraphics[width=1.\columnwidth]{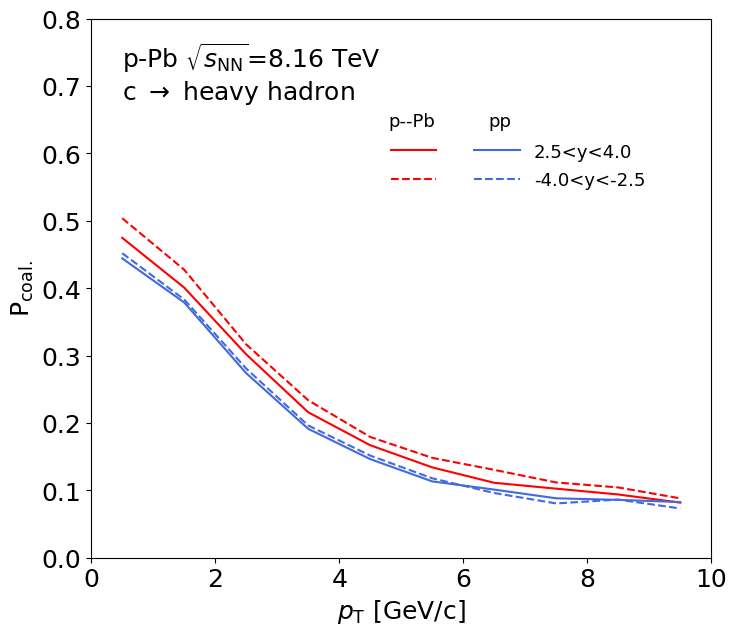}
\caption{(Color online) The coalescence probability of charm quarks at forward and backward rapidities to a charm hadrons, displayed as a function of the charm quark transverse momentum in pp and p--Pb collisions at $\sqrt{s_{\mathrm{NN}}}$ = 8.16 TeV.}
\label{Fig_Coal_Prob}
\end{center}
\end{figure}

In our previous work~\cite{Zhang:2024zga}, the implementation of independent fragmentation together with coalescence in the AMPT model was shown to be essential for describing the $D^0$ meson production in p--Pb collisions at 5.02 TeV. In this study, we further investigate the role of hadronization in describing the rapidity dependence of the $p_{\mathrm{T}}$ spectra and nuclear modification factor. To quantify the competition between coalescence and fragmentation, we first examine the charm quark coalescence probability as a function of its transverse momentum in different collision systems and rapidity intervals, as shown in Fig.~\ref{Fig_Coal_Prob}. The coalescence probability decreases with increasing $p_{\mathrm{T}}$ in all cases, indicating that high-$p_{\mathrm{T}}$ charm quarks are more likely to fragment independently rather than recombine with light partons. In pp collisions, the probabilities in the forward and backward rapidity regions are almost identical, reflecting the symmetry of the collision system. In p--Pb collisions, the coalescence probability is higher than in pp across the entire $p_{\mathrm{T}}$ range, suggesting that the presence of the lead nucleus enhances the local parton density and promotes recombination. Moreover, the coalescence probability in the backward (lead-going) region is significantly larger than that in the forward (proton-going) region. This rapidity asymmetry can be attributed to the higher multiplicity and parton density on the lead-going side, which provides more opportunities for charm quarks to coalesce with light partons.  

\begin{figure}[!hbt]
\begin{center}
\includegraphics[width=1.\columnwidth]{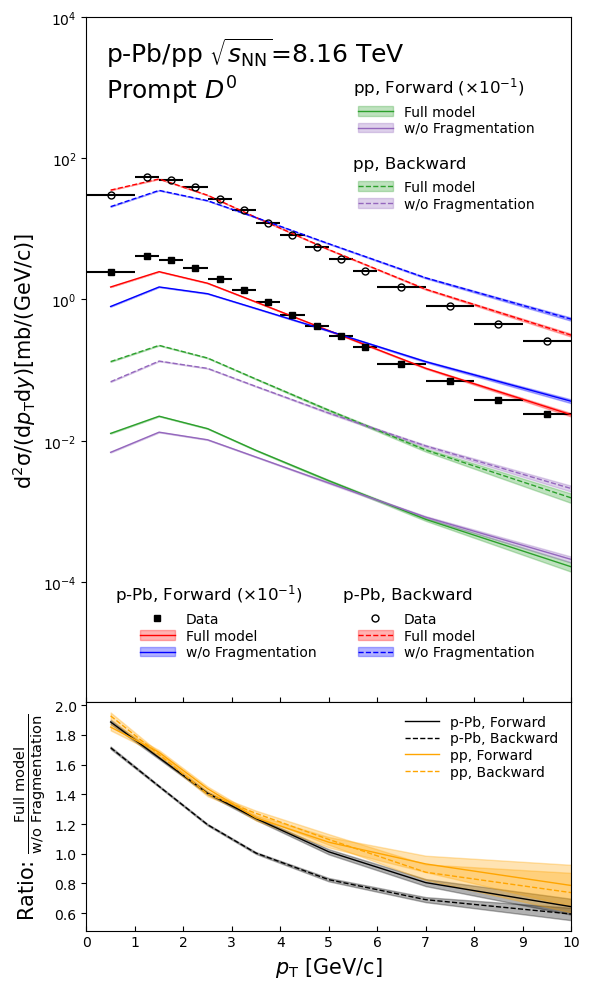}
\caption{(Color online) Double-differential cross-sections of prompt $D^0$ mesons in pp and p--Pb collisions at $\sqrt{s_\text{NN}}$ = 8.16 TeV. Top: spectra in forward (solid curves) and backward (dashed curves) rapidity intervals, for both systems with and without independent fragmentation. Bottom: ratio of spectra with independent fragmentation turned on to those with it turned off for the corresponding rapidity and collision system. The data points are from the LHCb Collaboration~\cite{LHCb:2022dmh}.}
\label{Fig_Spec_Frag}
\end{center}
\end{figure}

In Fig.~\ref{Fig_Spec_Frag}, we compare the $p_{\mathrm{T}}$ spectra of $D^0$ mesons with and without the fragmentation process in both $pp$ and $p$--Pb collisions at forward and backward rapidities. In all cases, enabling fragmentation (i.e. "Full model") enhances the $D^0$ yield at low $p_{\mathrm{T}}$ and suppresses it at high $p_{\mathrm{T}}$. This behavior reflects the kinematic redistribution of charm quarks from higher to lower $p_{\mathrm{T}}$ by the hadronization process, which depends on the coalescence probability as shown in Fig.~\ref{Fig_Coal_Prob}. When fragmentation is switched off, the charm quarks are forced to coalesce, and a coalesced $D$ meson retains almost the full charm-quark momentum, whereas fragmentation shares this momentum with the accompanying light hadrons, thereby shifting charm quarks to lower $p_{\mathrm{T}}$, leading to a crossing point between the two calculations. The location of this crossing point exhibits a clear dependence on the collision system and rapidity. For $pp$ collisions, the crossing point appears at approximately $5.5$ GeV/$c$ in both forward and backward rapidities, and the similar crossing point is also observed for $p$--Pb collisions in the forward region. In contrast, for $p$--Pb collisions in the backward region, the crossing point shifts to a lower $p_{\mathrm{T}}$ of about $3.5$ GeV/$c$, indicating that the enhanced coalescence probability in this region reduces the kinematic range over which fragmentation dominates. The lower panel of Fig.~\ref{Fig_Spec_Frag} shows the ratio of the $p_{\mathrm{T}}$ spectra with fragmentation to those without, providing a quantitative comparison of the fragmentation contribution. In $pp$ collisions, the ratios in the forward and backward rapidities are nearly identical. This confirms that fragmentation exhibits no rapidity dependence in a symmetric system. In $p$--Pb collisions, the ratio in the backward region deviates significantly from the $pp$ case, reflecting the stronger contribution of coalescence in this region. The forward–backward asymmetry of the ratio in p--Pb collisions can be understood from two effects: first, without fragmentation, the $D^0$ fraction is more suppressed in the forward region because the scarcity of light partons forces more charm quarks into hidden charm mesons instead of open charm hadrons; second, when fragmentation is turned on, these hidden charm states are dissolved and the released charm quarks hadronize into $D^0$, restoring the $D^0$ fraction to a similar level in both regions. Since the forward $D^0$ yield starts from a lower baseline, the same absolute increase leads to a larger relative enhancement. At low $p_{\mathrm{T}}$ ($\lesssim 4$ GeV/$c$), the ratio closely follows the $pp$ results, indicating that in the forward region, the fragmentation contribution remains close to that of a free $pp$ system without significant nuclear modification. At higher $p_{\mathrm{T}}$, the ratio gradually approaches that of the backward region. This suggests that even at high $p_{\mathrm{T}}$, the enhanced parton density in the nuclear environment begins to influence the hadronization balance. 

\begin{figure*}[!hbt]
\begin{center}
\includegraphics[width=2.\columnwidth]{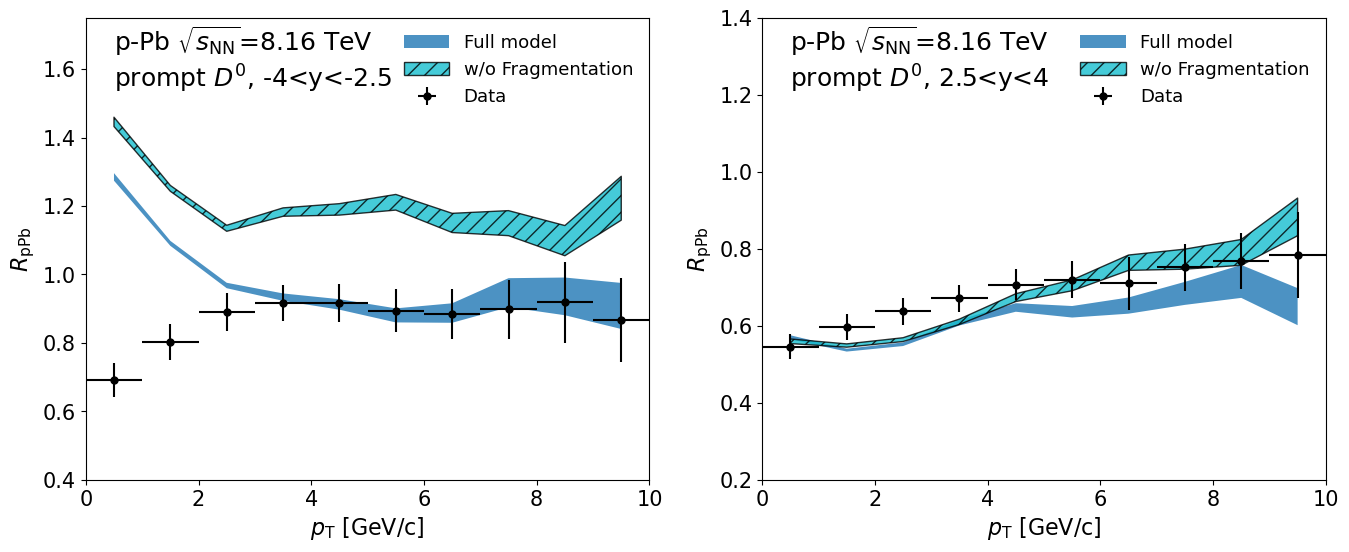}
\caption{(Color online) Nuclear modification factor as a function of $p_\mathrm{T}$ for prompt $D^0$ mesons in the (left) backward and (right) forward regions with and without independent fragmentation. The data points are from the LHCb Collaboration~\cite{LHCb:2022dmh}.}
\label{Fig_RpPb_Frag}
\end{center}
\end{figure*}

In Fig.~\ref{Fig_RpPb_Frag}, we examine how the hadronization mechanism influences the nuclear modification factor $R_{\mathrm{pPb}}$. The left panel shows $R_{\mathrm{pPb}}$ as a function of $p_{\mathrm{T}}$ for $D^0$ mesons in the backward region with and without fragmentation, together with the LHCb data~\cite{LHCb:2022dmh}, and the right panel displays the same calculation for the forward region. In the backward region, turning off fragmentation leads to a significant enhancement of $R_{\mathrm{pPb}}$ across the full $p_{\mathrm{T}}$ range. This indicates that with coalescence alone, the high parton density on the lead-going side would enhance $D^0$ production relative to $pp$. The inclusion of fragmentation dilutes this enhancement, because the more abundant coalescence in $p$Pb collisions reduces the number of charm quarks available for fragmentation compared to $pp$, thereby reducing the $R_{\mathrm{pPb}}$. In the forward region, the effect is $p_{\mathrm{T}}$-dependent. At low $p_{\mathrm{T}}$ ($\lesssim 4$ GeV/$c$), $R_{\mathrm{pPb}}$ almost unchanged when fragmentation is turned off. It is consistent with the finding from Fig.~\ref{Fig_Spec_Frag} that the fragmentation contribution in this kinematic range is similar to that in a free $pp$ system. At higher $p_{\mathrm{T}}$, $R_{\mathrm{pPb}}$ gradually increases as fragmentation is removed. This behavior reflects that in the nuclear environment, coalescence provides a small but non-negligible contribution to high-$p_{\mathrm{T}}$ production, which becomes relatively more important when fragmentation is turned off. These observations demonstrate that in addition to initial-state effects and partonic interactions, a proper description of $R_{\mathrm{pPb}}$ relies on a comprehensive treatment of hadronization, particularly the balance between coalescence and fragmentation across different kinematic regions.

\subsection{Anisotropic flow of prompt $D^0$ meson}

\begin{figure*}[!hbt]
\begin{center}
\includegraphics[width=1.33\columnwidth]{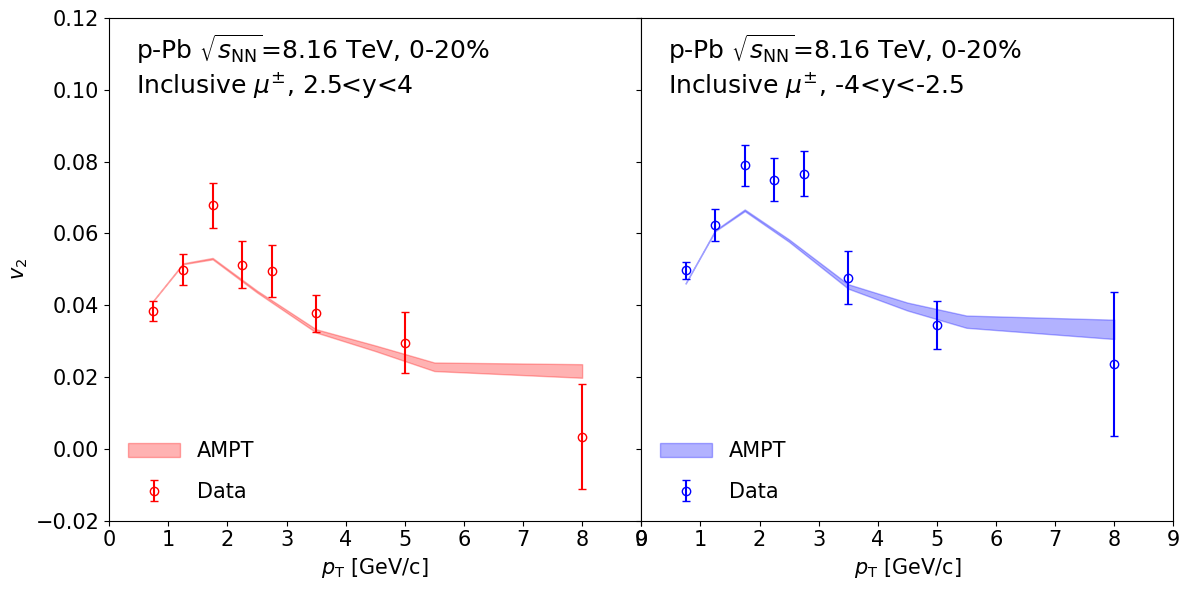}
\includegraphics[width=0.67\columnwidth]{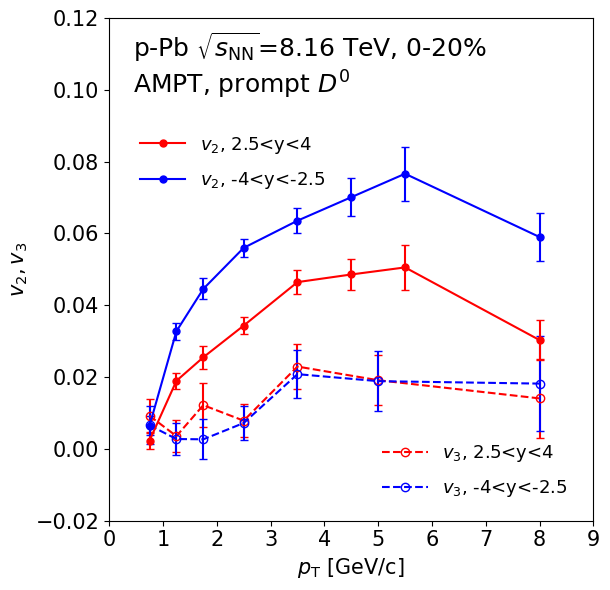}
\caption{(Color online) $p_{\mathrm{T}}$-differential $v_2$ and $v_3$ of prompt $D^0$ mesons and inclusive muons in high multiplicity(0-–20\%) p--Pb collisions at $\sqrt{s_{\mathrm{NN}}}$ = 8.16 TeV. Left: inclusive muon $v_2$ in forward and backward rapidity, respectively. The data points are from the ALICE Collaboration~\cite{ALICE:2022ruh}. Right: $D^0$ $v_2$ and $v_3$ in forward and backward rapidity intervals obtained from the AMPT calculations.}
\label{Fig3_D0v2}
\end{center}
\end{figure*}

Figure~\ref{Fig3_D0v2} presents the $p_{\mathrm{T}}$-differential $v_2$ and $v_3$ of prompt $D^0$ mesons and inclusive muons in high-multiplicity (0–20\%) p--Pb collisions at $\sqrt{s_{\mathrm{NN}}} = 8.16$ TeV from the AMPT model. Since no direct experimental measurement of $D^0$ $v_2$ is available in this collision system, we compare the model results with the muon $v_2$ data from the ALICE Collaboration~\cite{ALICE:2022ruh}. To obtain the muon $v_2$ in the model, we first compute the $v_2$ and yields of $D$ mesons, $B$ mesons, and light-flavor hadrons, then decay them to muons using \textsc{Pythia}8, following the same approach as in Ref.~\cite{ALICE:2022ruh}. The contribution from baryons(e.g. $\Lambda_{c}^{+}$) decay is not included due to its smaller production yield and lower semileptonic branching fraction relative to $D$ mesons. The model qualitatively reproduces the non-zero $v_2$ and its increasing trend with $p_{\mathrm{T}}$, and yields magnitudes consistent with the data, as shown in the left panel. This agreement indicates that the current model framework provides a reasonable description of heavy-quark collectivity in p--Pb collisions at forward and backward rapidities. In the right panel, the $D^0$ $v_2$ in the backward rapidity is significantly larger than that in the forward rapidity across the entire $p_{\mathrm{T}}$ range. 
The $D^0$ $v_3$ in the same panel is smaller than $v_2$ and exhibits no significant rapidity dependence, which is consistent with the expectation that $v_3$ is primarily driven by initial-state fluctuations. This observation establishes a reliable baseline for further investigations into the rapidity dependence of heavy-flavor collectivity and its response to partonic interactions and hadronization mechanisms in small collision systems.

\begin{figure*}[!hbt]
\begin{center}
\includegraphics[width=2.\columnwidth]{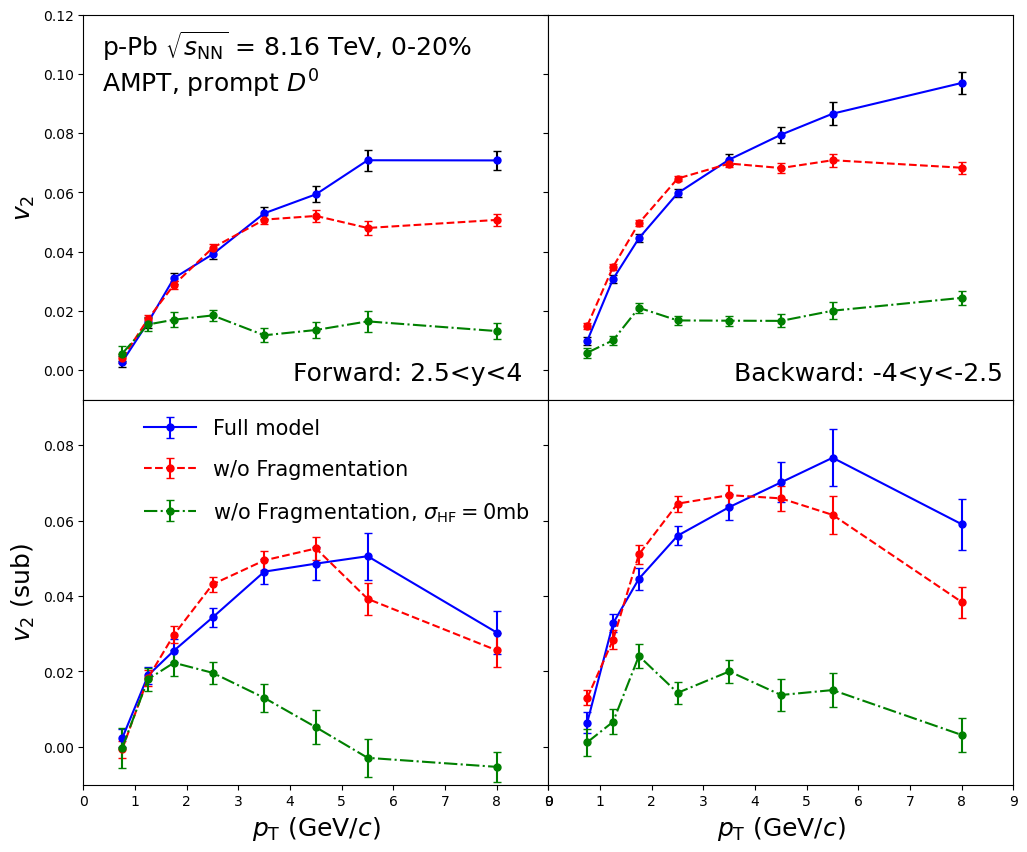}
\caption{(Color online) Comparison of $v_2$ of prompt $D^0$ mesons obtained from the AMPT model in forward (left) and backward (right) rapidity regions. The top panels show the results before the subtraction of long-range jet correlations, and the bottom panels show $v_2\{\mathrm{sub}\}$ after the subtraction. Different model configurations are compared: the full model (solid line), without fragmentation (dashed line), and without both fragmentation and heavy quark scattering (dot-dashed line, $\sigma_{\mathrm{HQ}} = 0$).}
\label{Fig_v2_Frag_Comp}
\end{center}
\end{figure*}

We now investigate how the hadronization mechanism and parton interactions affect the elliptic flow $v_2$ of $D^0$ mesons. Figure~\ref{Fig_v2_Frag_Comp} presents the $p_{\mathrm{T}}$-differential $v_2$ in forward and backward rapidities under three model configurations: the full model (solid line), without fragmentation (dashed line), and without both fragmentation and heavy quark partonic scatterings (dot-dashed line, $\sigma_{\mathrm{HQ}} = 0$). The upper panels show the results before the subtraction of long-range non-flow correlations, while the lower panels show the corresponding $v_2(\mathrm{sub})$ after subtraction. In the upper panels, a clear enhancement of $v_2$ is observed at high $p_{\mathrm{T}}$ ($> 4$ GeV/$c$) in both rapidity regions when fragmentation is included. This enhancement arises because fragmentation products originate from hard-scattered partons, and the resulting jet-like correlations contribute significantly to the measured $v_2$ before non-flow subtraction. In the lower pannels, the $v_2$ is significantly reduced after subtraction non-flow correlations in all cases and exhibits distinct $p_{\mathrm{T}}$-dependent patterns. At low $p_{\mathrm{T}}$ ($\lesssim 4$ GeV/$c$), turning off fragmentation leads to a larger $v_2$ compared to the full model calculation. This reflects that high-$p_{\mathrm{T}}$ charm quarks that fragment into low-$p_{\mathrm{T}}$ $D^0$ mesons carry weaker collective flow, thereby diluting the $v_2$ generated by coalescence. At high $p_{\mathrm{T}}$, the $v_2$ becomes larger when fragmentation is present, since fragmentation is the primary mechanism that converts the anisotropic motion of high-$p_{\mathrm{T}}$ charm quarks into the observed hadron $v_2$. Without fragmentation, high-$p_{\mathrm{T}}$ production relies on the much rarer coalescence process, which is less effective in conveying parton-level collectivity, resulting in a smaller $v_2$. When both fragmentation and heavy quark scatterings are disabled (dot-dashed line), the $v_2$ is significantly reduced across the entire $p_{\mathrm{T}}$ range and exhibits a negligible dependence on rapidity. It demonstrates that the non-zero $v_2$ seen in the full model is predominantly generated by scatterings of charm quarks with the medium during the parton cascade stage. Note that, even with $\sigma_{\mathrm{HQ}}=0$, the $v_2$ does not vanish completely but retains a small positive value, particularly at intermediate $p_{\mathrm{T}}$. This residual $v_2$ originates from light quarks, whose anisotropic flow—generated by parton scatterings and the parton escape mechanism—is transferred to charm hadrons via coalescence. 
Furthermore, the vanishing rapidity dependence in this configuration confirms that the larger $v_2$ in the backward region observed in full model is due to the higher parton density on the lead-going side, which enhances heavy quark rescattering. These results highlight that the combination of final-state partonic interactions and the competition between coalescence and fragmentation is essential for a comprehensive understanding of heavy-flavor $v_2$ across different rapidity regions in asymmetric collision systems. 

\begin{figure}[!hbt]
\begin{center}
\includegraphics[width=1.\columnwidth]{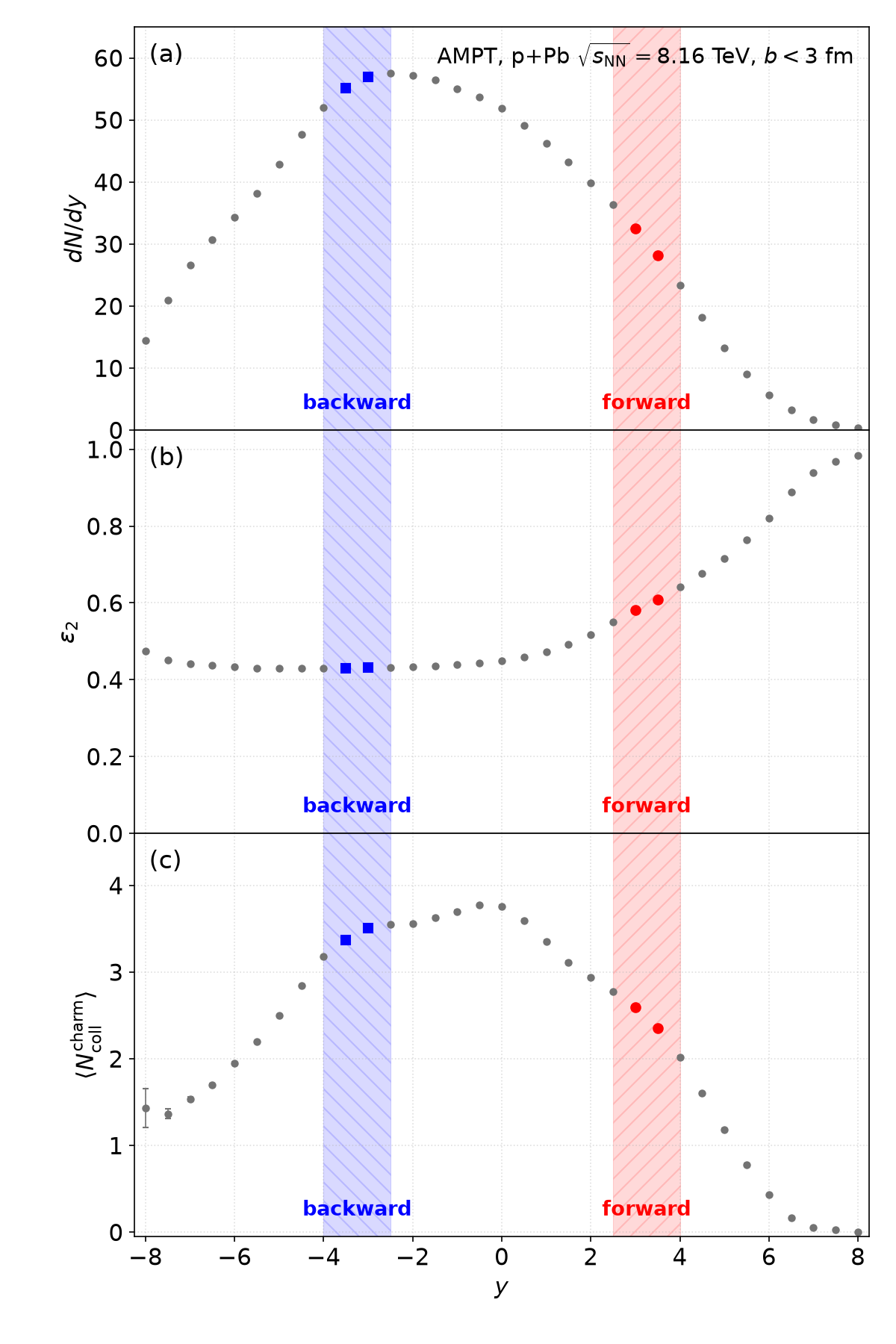}
\caption{(Color online) Rapidity dependence of the parton multiplicity $dN/dy$ (top), the initial-state eccentricity $\varepsilon_2$ (middle), and the average number of collisions per charm quark $\langle N_{\mathrm{coll}}^{\mathrm{charm}}\rangle$ (bottom) in p--Pb collisions at $\sqrt{s_{\mathrm{NN}}} = 8.16$ TeV with $b<3$ fm. The shaded regions mark the forward ($2.5<y<4$, proton-going) and backward ($-4<y<-2.5$, lead-going) rapidity windows.}
\label{Fig_e2_dndy_ncoll}
\end{center}
\end{figure}

To further quantify this final-state origin, we examine in Fig.~\ref{Fig_e2_dndy_ncoll} three rapidity-dependent quantities computed in the similar high-multiplicity (near-central, $b<3$ fm) sample as the $v_2$ measurement: the parton multiplicity $dN/dy$, the initial-state eccentricity $\varepsilon_2$, and the average number of collisions per charm quark, $\langle N_{\mathrm{coll}}^{\mathrm{charm}}\rangle$, defined as the total number of collisions involving charm quarks divided by the number of initial charm quarks. Here $\varepsilon_2$ is obtained from the transverse positions of the initial partons in each rapidity bin,
\begin{equation}
\varepsilon_2 = \frac{\sqrt{\big(\langle x^2\rangle-\langle y^2\rangle\big)^2 + 4\langle xy\rangle^2}}{\langle x^2\rangle+\langle y^2\rangle},
\end{equation}
where the centroid of the partons in each bin is subtracted and the principal axis defines the reference plane~\cite{Ma:2016hkg}. In the forward and backward rapidity windows ($2.5<|y|<4$), $\varepsilon_2$ is larger in the forward direction than in the backward direction, whereas $dN/dy$ and $\langle N_{\mathrm{coll}}^{\mathrm{charm}}\rangle$ are both larger in the backward direction. The larger $\varepsilon_2$ on the proton-going side is mainly dominated by fluctuations due to the smaller parton multiplicity there. The rapidity dependence of $\varepsilon_2$ is opposite to that of $v_2$, while $dN/dy$, $\langle N_{\mathrm{coll}}^{\mathrm{charm}}\rangle$, and $v_2$ all follow the same rapidity trend. The forward--backward asymmetry of $v_2$ thus originates from the different underlying parton densities on the proton-going and lead-going sides, which lead to different heavy-quark rescattering rates.

\section{Summary}
\label{sec:summary}
In this work, we have employed the heavy-flavor improved string-melting version of the AMPT model to perform a systematic study of prompt $D^0$ meson production and elliptic flow $v_2$ in p--Pb collisions at $\sqrt{s_{\mathrm{NN}}}=8.16$~TeV, and the forward--backward asymmetries observed in both observables. The model simultaneously describes the $R_{\mathrm{pPb}}$ and $v_2$ of heavy-flavor hadrons in both forward and backward rapidities. We found that this forward--backward asymmetry arises from the interplay of initial-state cold nuclear matter effects, final-state partonic interactions, and the competition between coalescence and fragmentation. In the forward region, nuclear shadowing suppresses $R_{\mathrm{pPb}}$ and the lower parton density reduces charm-quark rescattering, leading to a smaller $v_2$. In the backward region, anti-shadowing and the Cronin effect enhance $R_{\mathrm{pPb}}$ at low to intermediate $p_{\mathrm{T}}$, while the higher parton density promotes both coalescence and charm-quark rescattering, resulting in a larger $v_2$. The coalescence mechanism is more significant at lower $p_{\mathrm{T}}$ and in the backward region, whereas fragmentation becomes increasingly important at higher $p_{\mathrm{T}}$ and in the forward region. These results highlight the necessity of a unified framework that incorporates both initial- and final-state effects as well as proper heavy-quark hadronization to describe heavy-flavor observables across rapidity in small collision systems. Future measurements of additional heavy-flavor hadron species with higher precision at forward and backward rapidities will provide further constraints on these mechanisms.

\section{Acknowledgement}
The computation is completed in the HPC Platform of Wuhan Textile University. This work is supported in part by the National Natural Science Foundation of China (NSFC) under Grants No.~12405158 (S.T.), No.~12405159 (C.Z.), No.~12147101, No.~12547102 and No.~12325507 (G.-L.M.), the Natural Science Foundation of Hubei Province under Grant No.~2024AFB136 (C.Z.), and the U.S. National Science Foundation under Grant No.~2310021 (Z.-W.L.).

\bibliography{reference}
\end{document}